\documentclass{amsproc}
\usepackage{epsf}
\begin{document}
\newcommand{\identity}{\:\mbox{\sf 1} \hspace{-0.37em} \mbox{\sf 1}\,}
\newcommand{\ket}[1]{| #1 \rangle}
\newcommand{\bra}[1]{\langle #1 |}
\newcommand{\braket}[2]{\langle #1 | #2 \rangle}
\newcommand{\tr}{\rm tr}
\newcommand{\rank}{\rm rank}
\newcommand{\proj}[1]{| #1\rangle\!\langle #1 |}


\author{Elitza N. Maneva$^*$}
\address{$^*$ California Institute of Technology\\Pasadena CA 91126-0671}
\email{elitza@its.caltech.edu}
\author{John A. Smolin$^\dag$}
\address{$^\dag$ IBM T.J. Watson Research Center\\Yorktown Heights\\NY 10598}
\email{smolin@watson.ibm.com}
\thanks{The authors would like to thank Peter Shor for helpful discussions, and
the Summer Undergraduate Research Fellowship program at the California
Institute of Technology and IBM for support.  J.A. Smolin also thanks the
Army Research office for support under contract number DAAG55-98-C-0041.}

\title{Improved two-party and multi-party purification protocols}

\date{\today}

\begin{abstract}
We present an improved protocol for entanglement purification of 
bipartite mixed states using several states at a time rather
than two at a time as in the traditional recurrence method.  We
also present a generalization of the hashing method to n-partite
cat states, which achieves a finite yield of pure cat states for
any desired fidelity.  Our results are compared to previous
protocols.

\end{abstract}
\maketitle

%

\section{Introduction}
Entanglement is a fundamental resource in quantum information.  It can
be used for secure quantum cryptography \cite{ekert} 
and is an essential part of
known algorithms for quantum computation \cite{shor,grover}
(strangely, it is not known that all quantum algorithms which
outperform their classical counterparts {\em require} entanglement.
See \cite{qne} for a situation in which quantum states display a form
of nonlocality, but which involves no entanglement).

Early studies of entanglement purification \cite{purification,bdsw,qpa}
focused mainly on bipartite entanglement, attempting to distill pure
EPR pairs \cite{epr} from bipartite mixed states.  More recently,
Murao, Plenio, Popescu, Vedral and Knight \cite{mppvk} have studied
the generalization of such schemes to distilling three-party (GHZ \cite{GHZ}) 
and multi-party states of the form (sometimes called ``cat'' states
\cite{pawnote}) 
\begin{equation} 
\ket{\Phi^+}=\frac{1}{\sqrt{2}} (\ket{00\ldots 0} + \ket{11\ldots 1}) 
\end{equation} 
from three-party and multi-party entangled mixed states.
However, they do not study generalizations of
the hashing method of \cite{bdsw}.  The hashing scheme has the major
advantage over the recurrence style scheme of \cite{mppvk}
that it achieves a finite yield of pure cat states for any arbitrarily 
high fidelity, whereas the yield for any recurrence method goes to zero.

In this paper we study an improved purification protocol for two parties,
and a generalization of hashing to multiple parties.  

First, we define some notation:  All of our studies will apply to 
entangled mixed qubit states of $N$ parties (conventionally 
known as Alice, Bob, etc.), diagonal in the following
basis:
\begin{equation}
\ket{\psi_{p,i_1i_2\ldots i_{N-1}}}=\frac{\ket{0i_1i_2\ldots i_{N-1}}
+ (-1)^p \ket{1\bar{i}_1\bar{i}_2\ldots \bar{i}_{N-1}}}{\sqrt{2}}
\label{definepsi}
\end{equation}
where $p$ and the $i$'s are zero or one, and a bar
over a bit value indicates its logical negation.  This gives
$2^N$ orthogonal states.

These states correspond to the simultaneous eigenvectors of the following 
operators
(There are $N$ operators in all, one special one of the $X$ form, and
$N-1$ involving $Z$ and $I$):
\begin{equation}
\begin{array}{rccccccccc}
\\
S_0=&X&\otimes&X&\otimes&X&\otimes&X&\ldots&X\\
S_1=&Z&\otimes&Z&\otimes&I&\otimes&I&\ldots&I\\
S_2=&Z&\otimes&I&\otimes&Z&\otimes&I&\ldots&I\\
S_3=&Z&\otimes&I&\otimes&I&\otimes&Z&\ldots&I\\
&&&&&.\\
&&&&&.\\
&&&&&.\\
S_{N-1}=&Z&\otimes&I&\otimes&I&\otimes&I&\ldots&Z\\
\end{array}
\label{stabilizers}
\end{equation}
The $X$, $Z$, and $I$ operators, along with the $Y$ operator which we don't
use here, are members of the Pauli group (for more details, 
see~\cite{gotts}). 
The $p$ from Equation (\ref{definepsi}) corresponds to whether a state is a $+1$ or $-1$ eigenvector
of $S_0$ ($p=0$ for a $+1$ eigenvector and $p=1$ for a $-1$ eigenvector).  This
is called the ``phase'' bit of the cat state.
The $i_j$s correspond to whether a state is a $+1$ or $-1$ eigenvector of $S_j$ for 
$j=1,\ldots ,(N-1)$, which we call the amplitude bits.  
Thus, the following is the set of generators of the stabilizer group of
$\ket{\psi_{p,i_1i_2\ldots i_{N-1}}}$:
\begin{equation}
\{ (-1)^p S_0, (-1)^{i_1} S_1, \ldots, (-1)^{i_{N-1}} S_{N-1} \} 
\end{equation}
It is important to realize at this point that since these operators
are all tensor products of operators on the subsystems $1\ldots N$
each with eigenvalue $\pm 1$, they can be measured using only local
quantum operations plus classical communication. 
If an unknown one of the $\psi$'s is shared among $N$ parties and they
wish to determine the eigenvalue corresponding to one of the $S_i$,
each party just measures his or her operator and reports the result to 
everyone else.  The eigenvalue of the whole
operator is the product of their individual results.  Furthermore,
since $Z$ and $I$ commute, it is possible to measure the eigenvalues
of {\em all} the $S_{i>0}$.  On the other hand, $X$ and $Z$ do not
commute and therefore if $S_0$ is measured, none of the $S_{i>0}$ can
be measured (a random result would occur) and similarly if any of the
$S_{i>0}$ are measured the result of an $S_0$ measurement will be
randomized.  In other words, the parties can locally measure 
either all the amplitude bits or the phase bit for an unknown cat state.

\begin{figure}[htbf]
\centering
\epsfxsize=4in
\leavevmode\epsfbox{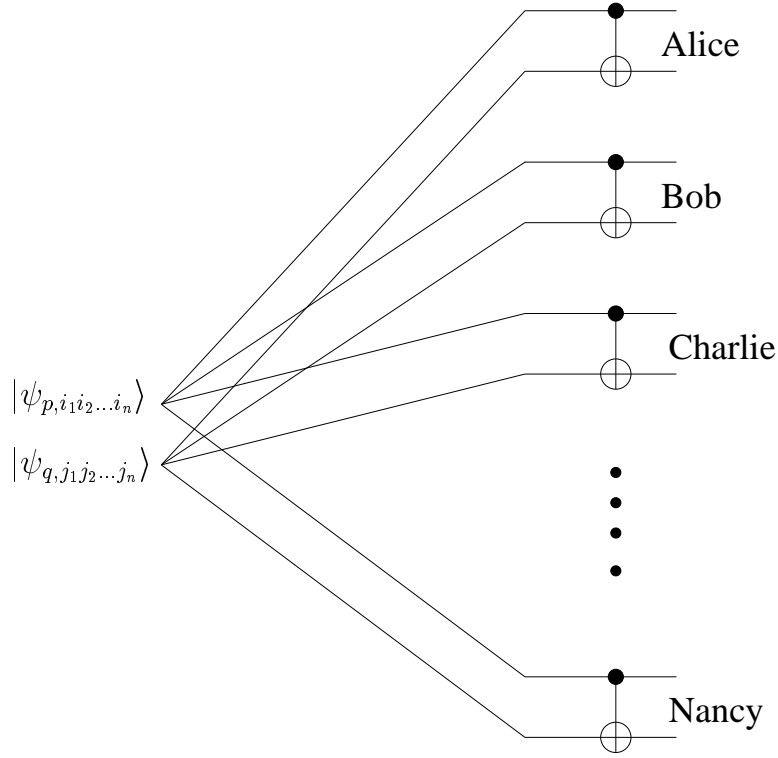}
\caption{The multi-party XOR.}
\label{multixor}
\end{figure}

The other tool we will need is the multilateral quantum XOR gate, in which
each party's bits are XORed together in a quantum-coherent way 
(see Fig. \ref{multixor}). 
Following Gottesman \cite{gotts} we can work out how a tensor product
of two cat states behaves under the multilateral XOR.  The 
generators of the stabilizer group behave as follows under the quantum
XOR operation:
\begin{equation}
\begin{array}{c}
X\otimes I \rightarrow X\otimes X\\
I\otimes X \rightarrow I\otimes X\\
Z\otimes I \rightarrow Z\otimes I\\
I\otimes Z \rightarrow Z\otimes Z\\
\end{array}
\label{xor}
\end{equation}

We work out here the case of three parties, the generalization
to $n$-partite cat states will be apparent.  Given $\ket{\psi_{p,i_1i_2}}$
and $\ket{\psi_{q,j_1j_2}}$ with stabilizers as in Eq. (\ref{stabilizers})
the generators of the stabilizers of the tensor product of 
these are given by:
\begin{equation}
\begin{array}{rrr}
\{ (-1)^p  XXXIII,& (-1)^{i_1} ZZIIII,& (-1)^{i_2}ZIZIII,\\
   (-1)^q IIIXXX,& (-1)^{j_1} IIIZZI,& (-1)^{j_2}IIIZIZ\}\\
\end{array}
\label{tensorform}
\end{equation}
(We have omitted the $\otimes$ symbol for brevity.)
Now, applying the rule for the XOR operation (\ref{xor}) to corresponding
operators (the first and fourth positions correspond to the first party's
piece of $\ket{\psi_{p,i_1i_2}}$ and $\ket{\psi_{q,j_1j_2}}$ respectively, 
etc.) we get:

\begin{equation}
\begin{array}{rrr}
\{ (-1)^p XXXXXX,& (-1)^{i_1} ZZIIII,& (-1)^{i_2}ZIZIII,\\
   (-1)^q IIIXXX,& (-1)^{j_1} ZZIZZI,& (-1)^{j_2}ZIZZIZ\}\\
\end{array}
\end{equation}

We can easily find another set of generators for the same
stabilizer group which is again the tensor form of (\ref{tensorform}):

\begin{equation}
\begin{array}{rrr}
\{ (-1)^{p+q} XXXIII,& (-1)^{i_1} ZZIIII,& (-1)^{i_2}ZIZIII,\\
   (-1)^q IIIXXX,& (-1)^{i_1+j_1} IIIZZI,& (-1)^{i_2+j_2}IIIZIZ\}\\
\end{array}
\end{equation}
This is simply the set of generators corresponding to
$\ket{\psi_{p\oplus q,i_1i_2}}\otimes\ket{\psi_{q,i_1\oplus j_1\ i_2\oplus j_2}}$.
What has happened is that the phase bits have been XORed together with 
the result put into the phase bit of the source state and the amplitudes
are each XORed together and stored in the target state's amplitude bits.
This suggests that the action of the multilateral quantum XOR gate (MXOR)
can be characterized by its action on the purely classical representation
of states as a set of bits, $(p,i_1,i_2,\ldots , i_{N-1})$:
\begin{eqnarray}
\lefteqn{{\rm MXOR}[(p,i_1,i_2,\ldots,i_{N-1}),(q,j_1,j_2,\ldots,j_{N-1})] =}\\
\nonumber
&&(p\oplus q, i_1,i_2,\ldots,i_{N-1}),
(q,i_1\oplus j_1,i_2\oplus j_2,\ldots,i_{N-1}\oplus j_{N-1})
\label{mxor}
\end{eqnarray}

Due to its linearity quantum mechanics allows us to think of mixed
states as if they are really one of the pure states in the mixture but
that we are simply lacking the knowledge of which one
(if the states in the mixture come with unequal probability, we are not
completely lacking knowledge of which state is in the mixture, but we only
know the probabilities, not which state we actually have).  Since all the
cat states (\ref{definepsi}) are interconvertible by local operations
\cite{isthisobvious} if we had a mixture of cat states and could
determine which one we actually had, we would be able to convert it to
a $\ket{\Phi^+}$ and would have purified the mixture. 

Putting everything we have said up to now together lets us find
purification schemes that are essentially classical; only the
rules of what we can do are given by quantum mechanics:
\begin{itemize}
\item Mixed states diagonal in the cat basis can be thought of as
being simply unknown members of the set of cat states.
\item The cat states (\ref{definepsi}) are all interconvertible by 
local operations, so determining which cat state one has is sufficient
to have purified it.
\item Either the $p$ or all the $i$'s of an unknown cat state may
be measured by local operations plus classical communication.
\item The multilateral XOR operation operates classically on the
$p$s and $i$'s of pairs of cat states according to Eq. (\ref{mxor}).
\end{itemize}
Our purification schemes will thus work by treating a set of many mixed states
(which are diagonal in the cat basis) as a set of unknown cat states,
and attempting to determine the unknown states, discarding them if we
cannot.  

We are now prepared to analyze the efficiencies of various
entanglement purification protocols applied to mixed states.
In particular, we concentrate on the generalization of the
Werner state \cite{werner}:
\begin{equation}
\rho_W=\alpha \proj{\Phi^+}  + \frac{1-\alpha}{2^N} \identity,
\ 0\le\alpha\le 1
\end{equation}
The fidelity of $\rho_W$ relative to the desired pure state
$\ket{\Phi^+}$ is $F=\bra{\Phi^+}\rho_W \ket{\Phi^+}=
\alpha+\frac{1-\alpha}{2^N}$.  We rewrite $\rho_W$ in the cat basis
(\ref{definepsi}) as 
\begin{eqnarray}
\lefteqn{\rho_W= (\alpha+\frac{1-\alpha}{2^N})
\proj{\psi_{0,00\ldots 0}}}\\
\nonumber && \mbox{} + \frac{1-\alpha}{2^N}\!\!\!\!\!\!\!\!\!\!\!\!\!\!\!
\sum_{p,i_1i_2\ldots i_{N-1} \ne 0,00\dots 0} \!\!\!\!\!\!\!\!\!\!\!
\!\!\!\!\!\!\proj{\psi_{p,i_1i_2\ldots i_{N-1}}}\ .  
\label{wernermixture}
\end{eqnarray}
Thus, the Werner state is diagonal in the cat basis
and we can think of it as really being one of the cat states.  We
write the unknown cat states as $N$ unknown strings of bits: $b_0,
b_1,b_2,\ldots, b_{N-1}$, where $b_0$ is formed by concatenating the
(unknown) phase bits of all the cat states, and the $b_j$ for $j>0$
are formed by concatenating the $j$th amplitude bits.  Together the
$b_j$ make up the total bitstring $B$.

\section{Bipartite Protocol}

The case of two parties has been studied \cite{purification,bdsw,qpa}.
The protocols can distill pure entanglement from any Werner state
with fidelity $F>1/2$.  The recurrence methods that work on
Werner states near $F=1/2$ involve local quantum operations on
two mixed states at a time.  For high fidelities, the best known strategy
(the {\em hashing} method) obtains high yields in the limit of arbitrarily
large numbers of states.  It seemed that operations on an intermediate number
of mixed states might give better yield for intermediate fidelities,
and this turns out to be the case.

\begin{figure}[htbf]
\centering
\epsfxsize=4in
\leavevmode\epsfbox{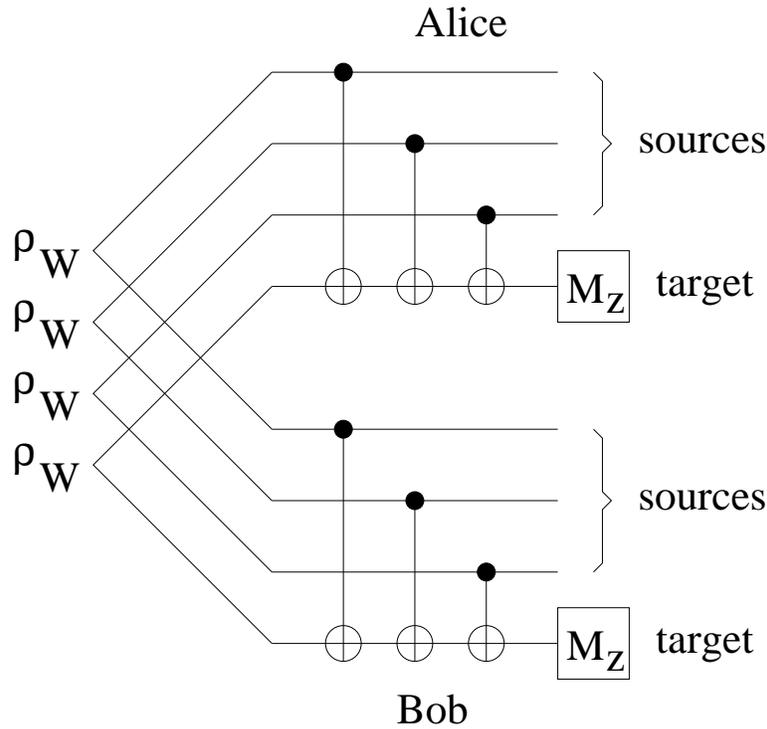}
\caption{The sequence of XOR gates and final measurement in the $z$
basis for our bipartite strategy for $m=4$.}
\label{nequals4}
\end{figure}

Our new strategy is to choose a block size $m$ and 
to take $m-1$ Werner states and do a bipartite XOR
between each one and an $m$th Werner state, and then to measure the
amplitude bits of that target state.  The sequence of XORs is
illustrated in Fig. \ref{nequals4} for the $m=4$ case.  This is a
natural generalization of the recurrence method whose single step is
just this method for $m=2$.  If any amplitude bit is nonzero, the
two measurements disagree, the source states are discarded.  If all
amplitude bits are zero the states are said to have ``passed'' and the
hashing method is performed on the $m-1$ source states along with
other source states that passed.  The advantage over the $m=2$
recurrence is that fewer than half the mixed states are used up
inherently just by being measured targets.

In \cite{bdsw} the hashing method was used only on states whose
mixture probabilities were {\em independent}.  We note that hashing
is a quite general method for extracting entropy from strings of
bits, even if there are correlations among the bits.  One merely
needs to take as many hash bits as there is entropy in the 
bitstring.  Thus, the yield of our method is:
\begin{equation}
p_{\rm pass} \frac{m-1}{m} 
\left(1-\frac{H({\rm passed\ source\ states})}{m-1}\right)
\end{equation}
The calculation of the entropy of a block of  passed states and of 
$p_{\rm pass}$ is straightforward.  One simply keeps track of the probability 
of each possible string $B$ (of $2 m N$ bits corresponding to a block of
$m$ Werner states) given the probabilities in the Werner mixture 
(\ref{wernermixture}), applies the MXOR rule (\ref{mxor}) to $B$ to yield
a $B'$ and groups the like $B'$s which have passed to yield
a final distribution $P_i$ (of the $2 (m-1) N$ bits corresponding $m-1$ states
left after the MXOR operation).  We then have $p_{\rm pass}=\sum_i P_i$
and the normalized distribution $P'_i=P_i/p_{\rm pass}$ and the 
entropy $H({\rm passed\ source\ states})$ given by $-\sum_i P'_i \log_2 P'_i$.

\begin{figure}
\centering
\epsfxsize=4in
\epsfbox{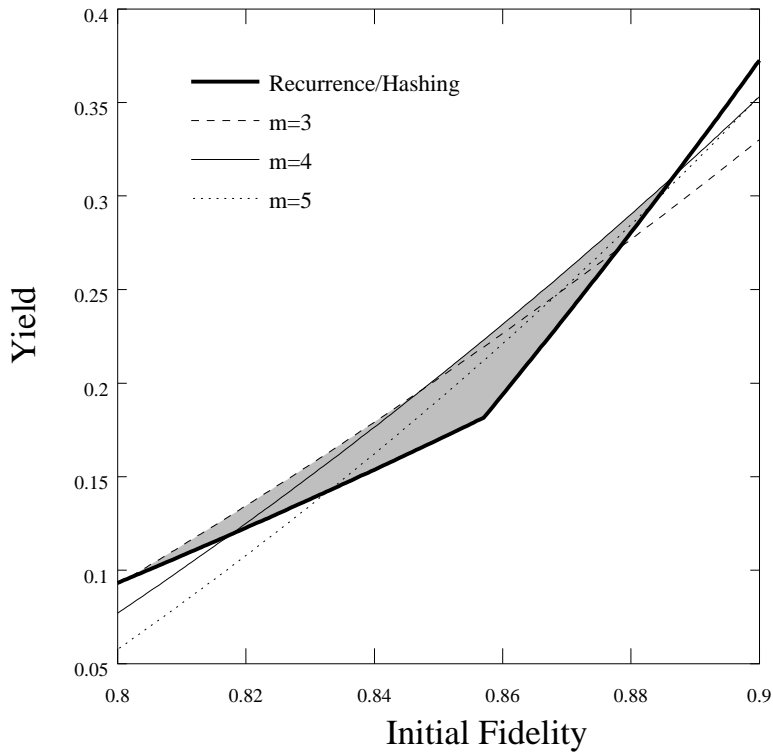}
\caption{Yields of various bipartite purification protocols.
The thick line is the recurrence continued by hashing method, the dashed line
is our new method for $m=3$, solid for $m=4$ and dotted for $m=5$.  The
shaded region is the region where our new method improves on the 
recurrence/hashing method.  Note that yield for $m=5$ is never the best,
though for some fidelities it is better than the recurrence/hashing 
method.  The $m=6$ case (not shown) behaves similarly, while for $m>6$
the new method is always worse than recurrence/hashing.
The sharp ``knee'' visible in the recurrence/hashing line
is the point above which recurrence is never used and hashing is done
immediately.}
\label{graph}
\end{figure}

Figure \ref{graph} compares the yield for our new method for various
values of $m$ with the previous recurrence continued by hashing
protocol.  We have not found a simple way to analyze what happens if
our multi-bit step is iterated, rather than passing on immediately to
hashing.  For the $m=2$ recurrence only one passed source state
remains and it is identical in all respects to every other passed
source state.  For $m>2$ there are multiple correlated passed states
and it is not clear just how to treat them.  For instance, at $m=3$
there are two passed states from each operation and there is no way to
combine the 4 passed states from two operations into another $m=3$
step.  

\section{Multipartite Hashing}
In \cite{mppvk} multi-party recurrence methods are studied, but not
multi-party hashing which is needed to achieve finite yields.  Here we
present a multi-party hashing method.

\begin{figure}[htbf]
\centering
\epsfxsize=5in
\leavevmode\epsfbox{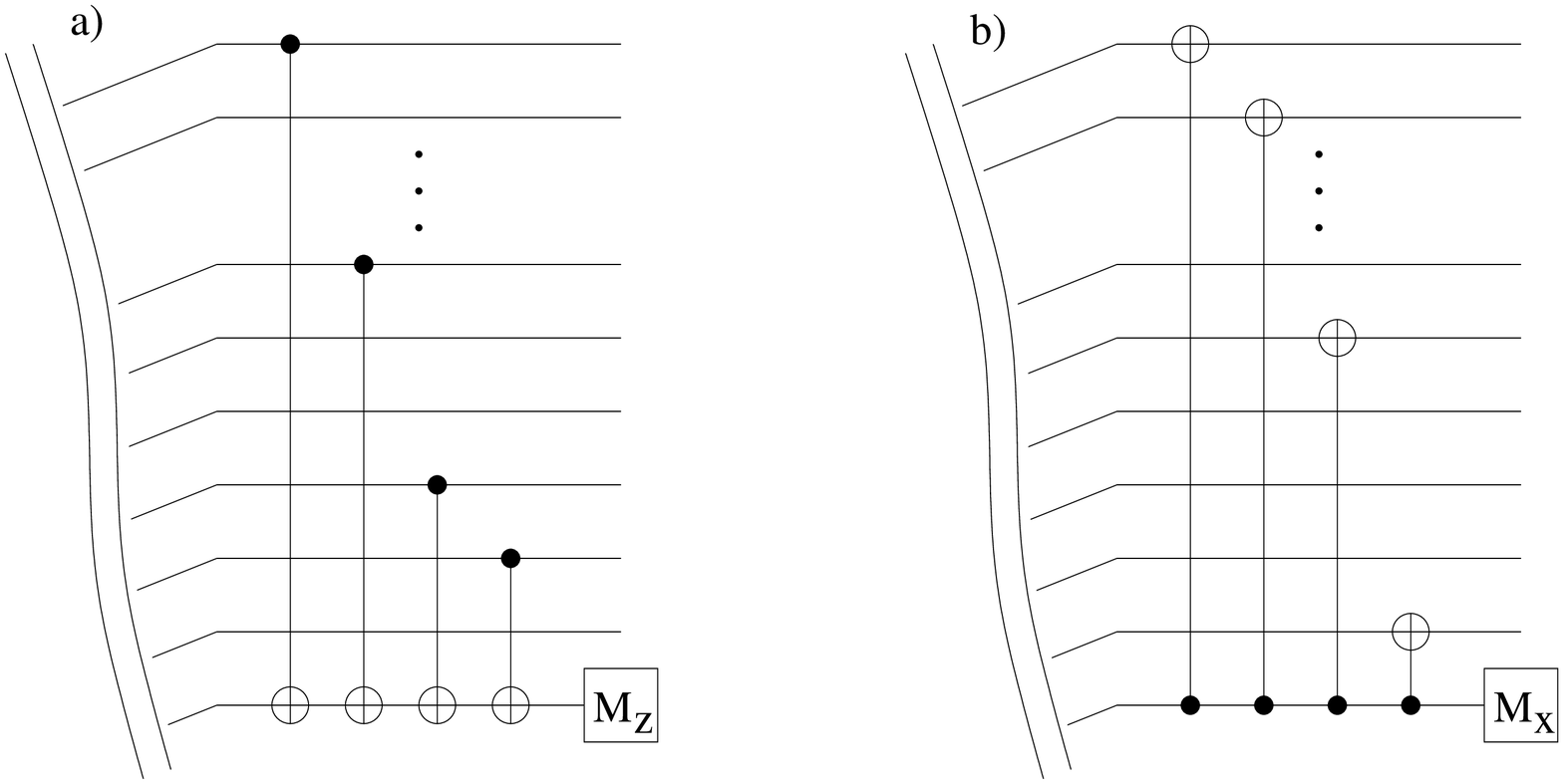}
\caption{Multi-party hashing: These hashes are done on large blocks of
bits (indicated by the vertical ellipsis) and are done multilaterally (only
one party's operations are shown, the other $N-1$ parties operations
are identical).\protect\\
a) Finding a random subset parity on 
all the $b_{j>0}$ in parallel.  In this case the first, third, sixth
and seventh states shown are XORed multilaterally into the last one
which is then measured to determine the eigenvalue of the $Z$ 
operator.\protect\\
b) Finding a random subset parity on
$b_0$.  In this instance the parity of the first, second, fourth
and eighth states shown are XORed with the last one, which 
is then measured in the eigenbasis of the $X$ operator.  Note the reversal 
of the direction of the XOR gates with respect to a).  
}
\label{multihash}
\end{figure}

In the case of two parties it is known \cite{bdsw} how to extract the
parity of any random subset of all the bits in $B$.  For more than two
parties it is not known now to do this.  Instead, we can choose to
extract any random subset parity on either the parity bitstring $b_0$
or on all the amplitude bitstrings $b_j$, $j>0$ in parallel.  This
follows immediately from Eq. (\ref{mxor}):  This is done  multilaterally
XORing together all the states in the desired subset choosing one of them
to be the target.  See Fig. \ref{multihash}.  Depending on the 
direction of the XOR gates either the phase bits or all the amplitude
bits accumulate in the target state, which can then be measured.

Our multilateral hashing protocol will be to choose a large block size
$m$ and then to extract $m \max_{j>0}[\{{H(b_j)}\}]$
random subsets of each amplitude bitstring in parallel (as shown
in Figure \ref{multihash}a), where $H(b_j)$ is
the entropy per bit in string $b_j$.  This is sufficient to determine
all the bits of all the $b_{j>0}$ as it is just doing the same
random hash on each bitstring.  Even though the random hashes are
all the same, since they are uncorrelated with the bitstrings
being determined this many hash bits will be enough to determine
all bits of the $b_{j>0}$ \cite{correlated}.  
This procedure actually extracts too much information (and thereby uses
up too many states as measured targets), so perhaps a more efficient 
protocol exists, but this has the virtue of using only the multilateral 
XOR operation
which maps cat states to cat states.  After determining the amplitude
bits, to find $b_0$ we use multilateral XORs arranged as in 
Fig. \ref{multihash}b, and find the hash of the
string by measuring another $H(b_0)$ of the states.  The yield of this
hashing protocol $D_h$ is given by
\begin{equation}
D_h=(1-\max_{j>0}[\{{H(b_j)}\}] -H(b_0))
\label{dh}
\end{equation}

\begin{figure}
\epsfxsize=4in
\center{\epsfbox{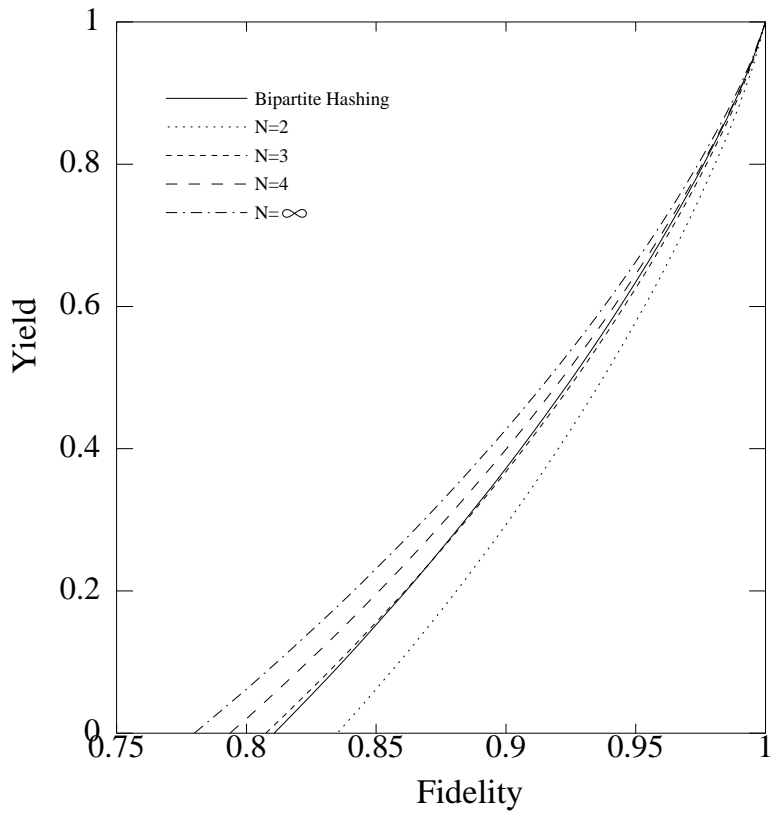}}
\caption{Yields for multipartite hashing for various numbers
of parties.  The solid line is the two-party hashing method of 
\protect\cite{bdsw}.  The dotted line is the corresponding $N=2$ 
version of our new hashing method, which has a lower yield since
it works on the amplitude and phase bits as separate hash strings
even though for the bipartite case it is known how to extract
their entropy together, which is more efficient.  The lines consisting
of dashes, longer dashes and dots with dashes are the $N=3$, $N=4$,
and $N=\infty$ cases respectively.
}
\label{graph2}
\end{figure}

For the case of Werner states all the $b$'s have the same entropy and
Eq. (\ref{dh}) reduces to
\begin{equation}
D_W=1-2 H_2\!\left(\frac{(1-f) 2^{N-1}}{2^N-1}\right) 
\label{dhw}
\end{equation}
or in the limit as the number of parties goes to infinity
\begin{equation}
D_W^\infty=1-2 H_2\!\left(\frac{1-f}{2}\right).
\end{equation}
where $H_2(x)=-x \log_2{x} - (1-x)\log_2{(1-x)}$.  
Eq. \ref{dhw} is graphed for several values of $N$ in Figure \ref{graph2}.
By using the recurrence method and switching to hashing as soon as it gives 
better yield, one can obtain positive final yield to arbitrarily high fidelity
for any initial fidelity for which the recurrence method of \cite{mppvk} 
improves the fidelity.

\section{Conclusions and Comments}

We have found improved bipartite recurrence protocols for 
the purification of entanglement from mixed quantum states.
We have also demonstrated the first finite-yield method 
for purification of cat states in a multi-party setting.
It is worthwhile to note that both of these new procedures
were analyzed for mixed state diagonal in the cat basis, but
that in fact they will work for {\em any} mixed state just
as well, by considering the state's cat-basis diagonal elements.
This is unlikely to be the optimally efficient strategy for
non-diagonal states however.  In \cite{bdsw} there is an
example of a state for which the conventional bipartite recurrence
and hashing cannot distill any pure entanglement, but which can
nevertheless simply be distilled.  One expects such examples to
exist for our new methods as well (indeed, the example in \cite{bdsw}
{\em is} an example for the bipartite case of  new methods which will 
similarly fail to distill it.

For our bipartite protocol, while clearly not optimal it is not so bad
to have passed over to hashing instead of recurring the protocol.
Recurrence methods have vanishing yield if one desires arbitrarily
high fidelity of the purified states, so hashing needs to be used
eventually in any case.  Additionally it would likely be best to
produce a variable block size protocol that begins as the recurrence
method for low fidelity, switches to a larger block size at some
higher fidelity and finally is continued by hashing.  A calculation of
the yield of such a method is cumbersome, and seemingly provides
little insight.  We hope that our having pointed out that block size
$m>2$ methods can improve over recurrence will stimulate further work
in this area to develop a deeper understanding, rather than just a
brute-force analysis.  Much progress has been made on purification
involving only one-way classical communication.  Such protocols
directly correspond to quantum error-correcting codes
(cf. \cite{bdsw}) but recurrence protocols inherently involve two-way
classical communication so all parties know which states to discard.
So far little of the coding theory has been applied to this case.
There does appear to be some relation between these two-way
purification protocols and quantum error-detecting codes, and some
progress is being made in this area \cite{gotts2}.


\begin{thebibliography}{99}
\bibitem{pawnote} These are known as cat states since they are
generalizations of the state of the many particles making up 
Schr\"oedinger's cat, namely 
$(\ket{{\rm alive\ alive\ alive}\ldots\ {\rm alive}}
+\ket{{\rm dead\ dead\ dead}\ldots\ {\rm dead}})/\sqrt{2}$.
\bibitem{correlated} If the bits of the various $b_{j>0}$ are 
uncorrelated for different values of $j$ this can be thought of
as taking the same random hash on different sets of data.  Since
the data are independent of the hash this will work just as well
as if different hashes were chosen for each $j$.  
In the case where the $b_{j>0}$ are correlated the situation can only get
better.  Consider the case of complete correlation--when all the bits in 
all the strings are the same the same hashes are clearly sufficient to
determine them all.
\end{thebibliography}
\end{document}